\documentclass[12pt,a4paper]{article}
\usepackage[german, english]{babel}
\usepackage{amsmath}
\usepackage{amssymb}
\usepackage{graphicx}
\usepackage{ifthen}
\usepackage{epsfig}
\usepackage{hyperref}
\newcounter{fig}   \newcommand{\lbfig}[1]{\refstepcounter{fig}
\label{#1} }

\newcommand{\Tr}{{\rm Tr}}

\newcommand{\bea}{\begin{eqnarray}}
\newcommand{\eea}{\end{eqnarray}}
\newcommand{\be}{\begin{equation}}
\newcommand{\ee}{\end{equation}}

\newcommand{\BU}{\ensuremath{\mathbf{U}}}
\newcommand{\BH}{\ensuremath{\mathbf{H}}}
\newcommand{\BR}{\ensuremath{\mathbf{R}}}

\newcommand{\Btau}{\ensuremath{\boldsymbol{\tau}}}

\newcommand{\Br}{\ensuremath{\mathbf{r}}}

\newcommand{\dd}{\mathrm{d}}
\newcommand{\ii}{\mathrm{i}}

\begin{document}

\title{Hopfions interaction from the viewpoint of the product ansatz}

\author{
{\large A.~Acus}$^{\dagger}$,
 {\large E.~Norvai\v{s}as}$^{\dagger}$
and {\large Ya. Shnir}$^{\star \ddagger}$ \\ \\
\\ $^{\dagger}${\small Vilnius University, Institute of Theoretical Physics and Astronomy}
\\ {\small Go\v{s}tauto 12, Vilnius 01108, Lithuania}
\\ $^{\star}${\small BLTP, JINR, Dubna, Russia}
\\ $^{\ddagger}${\small Institute of Physics, Carl von Ossietzky University Oldenburg, Germany}
} \maketitle

\begin{abstract}
We discuss the relation between the solutions of the Skyrme model of lower degrees and the corresponding axially
symmetric Hopfions which is given by the projection onto the coset space $SU(2)/U(1)$.
The interaction energy of the Hopfions is evaluated
directly from the product ansatz. Our results show that if the separation between
the constituents is not very small, the product ansatz can be
considered as a relatively good approximation to the general pattern of the
charge one Hopfions interaction both in repulsive and attractive channel.
\end{abstract}


\section{Introduction}

Spatially localized particle-like non-perturbative soliton field
configurations have a number of applications in a wide variety of
physical systems, from modern cosmology and quantum field theory
to condensed matter physics.
The study of the interaction between
the solitons and their dynamical properties has attracted a lot of
attention in many different contexts (for a general review see
e.g. \cite{Manton-Sutcliffe}). One of these interesting contexts include investigation of
a new family of materials known as topological
insulators, which also makes relevant the basis research involving
topological solitons. Perhaps the most interesting possibility is the discovery that frustrated magnetic materials
may support topological insulator phases, for which wave functions are classified by the Hopf invariant~\cite{Moore2008}.

Simple example of topological soliton solutions is given by the class of scalar
models from the Skyrme family, the original Skyrme model \cite{Skyrme:1961vq},
the Faddeev-Skyrme model \cite{Faddeev} in $d=3+1$, and the low-dimensional baby Skyrme model
in $2+1$ dimensions \cite{Bsk}. The Lagrangian of all these models as they were formulated originally,
has similar structure, it includes the usual sigma-model kinetic term,
the Skyrme term, which is quartic in derivatives, and the potential term which does not
contain the derivatives. According to the Derrick's theorem \cite{Derrick}, the latter term is optional
in $d=3+1$, however it is necessary to stabilise the soliton configurations in the baby-Skyrme model.

A peculiar feature of these models is that the corresponding
soliton solutions, Skyrmions and Hopfions, do not saturate the
topological bound. In order to attain the topological lower bound
and get a relation\footnote{This relation is linear for Skyrmions, however for the Hopfions
the Vakulenko-Kapitanski bound in $d=3+1$ is $E=c Q^{3/4}$ \cite{VK}
where $c$ is some constant. } between the masses of the solitons and
their topological charges $Q$, one has to modify the model, for
example drop out the quadratic kinetic term
\cite{Adam:2010fg,Foster:2010zb} or extend the model by coupling
of the Skyrmions to an infinite tower of vector mesons
\cite{Sutcliffe:2011ig}. Thus, the powerful methods of
differential geometry cannot be directly applied to describe
low-energy dynamics of the Skyrmions and Hopfions, one has to
analyse the processes of their scattering, radiation and
annihilation numerically \cite{Piette:1994mh,Battye:1996nt}.
Interestingly, the numerical simulations of the head-on collision
of the charge one Skyrmions reveal the celebrated picture of the
$\pi/2$ scattering through the intermediate axially-symmetric
charge two Skyrmion \cite{Battye:1996nt}, which is typical for BPS
configurations like vortices or monopoles (see
\cite{Manton-Sutcliffe}). The same pattern was observed in the
baby Skyrme model using the collective coordinate method
\cite{Sutcliffe:1991aua}. However recent attempt to model the
Hopfion dynamics \cite{Hietarinta:2011qk} failed to find the
channel of right-angle scattering in head-on collisions.

Typically, the problem of direct simulation of the soliton dynamics is related with sophisticated
numerical methods, the calculations require considerable amount of computational resources, actually this problem is
fully investigated only for the low-dimensional baby Skyrme model. Even more simple task of full numerical
investigation of the
spinning solitons beyond rigid body approximation was performed only recently in the Faddeev-Skyrme model
\cite{BattyMareike,JHSS} and in the baby Skyrme model \cite{Halavanau:2013vsa,Battye:2013tka}, in the case of
the original Skyrme model in $d=3+1$ this problem is not investigated yet.

Alternatively, one can delve into the assumptions about the character of the
soliton interaction by analogy with the dynamical
properties of the Bogomol'nyi type solitons \cite{Manton:1988ba,Sutcliffe:1991aua,Schroers:1993yk}. Then
the moduli space approximation for low-energy soliton dynamics can be applied.
This approach works especially well for low-dimensional baby Skyrme model because it can be considered
as a deformation of the $O(3)$ sigma model. It also explains the observations of the right-angle scattering in the
head-on collisions of the Skyrmions in $d=3+1$, however the question about validity of the moduli approximation to the
low-energy dynamics of the Hopfions is not quite clear.

Another approach to the problem of interaction between the solitons is to consider
the asymptotic field of the configurations, then for example the Skyrmions can be treated as triplets of scalar dipoles
\cite{Schroers:1993yk,Manton1994,Manton:2002pf}.
Similarly, the asymptotic fields both the  baby Skyrmion and the Hopfion in the sector of degree one correspond
to a doublet of orthogonal dipoles \cite{Piette:1994mh,Gladikowski:1996mb,Ward:2000qj}. Considering this system Ward
predicted existence of three attractive channels in the interaction of the charge one
Hopfions with different orientation \cite{Ward:2000qj}.
It was suggested recently to use
a simplified dipole-dipole picture of the interaction between the baby Skyrmions in the "easy plane" model, thus in this
description the interaction energy depends only on the average orientation of the dipoles \cite{Jaykka:2010bq}.

In his pioneering paper \cite{Skyrme:1961vq} Skyrme suggested to apply
the product ansatz which yields a good approximation to a configuration
of well-separated unit charge Skyrmions. The ansatz is constructed by the multiplication of
individual Skyrmion fields, besides the rational map ansatz \cite{Houghton:1997kg} it can be  used to
produce an initial multi-Skyrmion configuration for consequent numerical calculations \cite{Battye1998}.

In a similar way one can construct a system of well-separated baby-Skyrmions using
the parametrization of the scalar triplet in terms of the $SU(2)$-valued hermitian matrix fields \cite{Acus:2009df}.
Evidently, the same approach can be used to model the configuration of well separated static Hopfions of degree one.
On the other hand the product ansatz can be applied in the Faddeev-Skyrme model to approximate various multicomponent configurations
whose position curve consists of a few disjoint loops, like the $Q=4$ soliton.

In this Letter we discuss the relation between the solutions of the Skyrme model of lower degree and the corresponding axially
symmetric Hopfions which is given by the projection onto the coset space $SU(2)/U(1)$.
Using this approach we construct the product ansatz of two well-separated single Hopfion configurations.
We confirm that the product ansatz correctly reproduces the channels of interaction. Indeed, it is known that
similar with the case of the Skyrmions, the interaction between the two Hopfions can be repulsive or
attractive depending upon the relative orientation of the solitons  \cite{Ward:2000qj}.

\section{The model}
Let us consider a Faddeev-Skyrme model Lagrangian in 3+1 dimensions with metric $(+,-,-,-)$:
\begin{equation}
\label{model}
{\cal L} = \frac{1}{32\pi^2}\left(\partial_\mu \phi^a \partial^\mu \phi^a -
\frac{1}{4}(\varepsilon_{abc}\phi^a\partial_\mu \phi^b\partial_\nu \phi^c)^2 \right)\,.
\end{equation}
Here $\phi^a = (\phi^1, \phi^2,\phi^3)$ denotes a triplet of scalar real
fields which satisfy the constraint $|\phi^a|^2=1$.
The finite energy configurations should approach a constant value at spatial infinity, which
we selected to be $\phi^a(\infty) = (0,0,1)$. Thus, the static field $\mathbf{\phi}(\mathbf{x})$ defines
a map $R^3 \rightarrow S^2$, which can be characterized by Hopf invariant $Q = \pi_3(S^2) = \mathbb{Z}$.
Then the finite energy solutions of the
model, the Hopfions, are the map $S^3 \to S^2$ and the
target space $S^2$ by construction is the coset space $SU(2)/U(1)$.

It follows that any coset space element $\BH$ can be projected
from generic $SU(2)$ group element $\BU$. In circular coordinate system the projection can be written in the following form,
\begin{equation}
\label{genericProjection}
\BH=2\sum_a (-1)^a \tau_a \phi_{-a}= 2 \BU \tau_0 \BU^\dagger\,,
\end{equation}
where the Pauli matrices $(\tau_1,\tau_0,\tau_{-1})$ are chosen to satisfy relation
\begin{equation}
\label{pauliDefinition}
\tau_a \tau_b =\frac14 (-1)^a \delta_{a,-b}\mathbf{1} -\frac{1}{\sqrt{2}}
\left[
\begin{matrix}
1 & 1 &1\\
a & b & c
\end{matrix}
\right]\tau_c ,
\end{equation}
and $\left[
\begin{smallmatrix}
1 & 1 &1\\
a & b & c
\end{smallmatrix}
\right]$ denotes the Clebsch-Gordon coefficient.
It is convenient to rewrite the Lagrangian~\eqref{model} directly in terms of coset space elements $\BH$,
\begin{equation}
\label{modelInH}
{\cal L} = \frac{1}{64\pi^2}\left(\Tr\big\{\partial_\mu \BH \partial^\mu \BH\big\} +
\frac{1}{16}\Tr\big\{\bigl[\partial_\mu \BH,\partial_\nu \BH\bigr]\bigl[\partial^\mu \BH,\partial^\nu \BH\bigr]\big\} \right)\,.
\end{equation}

The difference between the Skyrmions and Hopfions is that in the latter case the dimensions of the domain space
and the target space are not the same, the topological charge of the Hopfions is not defined locally. It has a meaning of the linking
number in the domain space \cite{Faddeev}.

There have been many investigations of the solutions of the model~\eqref{model}
\cite{Gladikowski:1996mb,Sutcliffe:2007ui,Battye1998,Hietarinta2000}. Here we restrict out consideration to the axially symmetric
configurations of lower degrees $Q=1,2$ which are conventionally labeled as ${\cal A}_{1,1}$ and ${\cal A}_{2,1}$ \cite{Sutcliffe:2007ui}.

An approximation to these solutions can be constructed via Hopf
projection  of the corresponding Skyrmion configurations with baryon numbers
$B=1$ and $B=2$, respectively. Indeed, it have been shown~\cite{Battye1998,Su:2008} that up to
a constant, the solution for the charge $Q=1$ Hopfion  can be written in a form which is equivalent to the
standard hedgehog solution of the Skyrme model with the usual profile function $F(r)$. This construction yields the
hopfion with mass $1.232$. In our conventions ~\eqref{modelInH} the ${\cal A}_{1,1}$ configuration is $\BH_1$ which is a projection
of the Skyrmion matrix valued field $\BU_0$, i.e.
\begin{equation}
\label{hopfion1Projection}
\BH_1(\Br)=2\BU_0(\Br) \tau_0 \BU_0^\dagger(\Br)\,,
\end{equation}
where $\BU_0(\Br)$ denotes the usual spherically symmetric Skyrmion which is parametrised via the hedgehog ansatz
\begin{equation}
\label{skyrmionAnsatz}
\BU_0(\Br)=\exp \bigl(2\ii (\hat{\Br}\cdot\Btau) F(r)\bigr)\,.
\end{equation}
Here $\hat{\Br}$ denotes the unit position vector and $F(r)$ is a monotonically decreasing profile
function of the Skyrmion with usual boundary conditions, $F(0)=0,~~F(\infty)=\pi$.

In terms of the triplet scalar fields $\phi_a$ in the circular coordinate system defined by $\phi_{\pm 1}=\mp\frac{1}{\sqrt{2}}\bigl(\phi^1\pm\mathrm{i}\phi^2\bigr)$ and $\phi_0=\phi^3$, the projected
Skyrme configuration can be written as\footnote{Note there is a misprint in the corresponding expression for the component
$\phi_2$ in~\cite{Acus:2009df}}
\begin{equation}
\label{skyrmionAnsatzInphi}
\phi_a=2\sin^2 F(r) \hat{r}_0 \hat{r}_a +\mathrm{i} a \sin \bigl(2 F(r)\bigr) \hat{r}_a + \cos \bigl(2 F(r)\bigr) \delta_{0,a} \,.
\end{equation}
Evidently, although the ansatz ~\eqref{skyrmionAnsatz} depends on the radial variable only, expression \eqref{skyrmionAnsatzInphi}
clearly demonstrates, that the corresponding  ${\cal A}_{1,1}$ Hopfion does not possess the spherical symmetry, the
projection breaks it down to axial symmetry \cite{Battye1998}.

The residual $O(2)$ symmetry of global rotations
by the phase $\alpha$ around the third axis in the internal space changes the triplet in the following way
\begin{equation}
\label{hopfionrotation}
\phi_{+1} \rightarrow  \phi_{+1} \mathrm{e}^{i\alpha},\quad   \phi_{-1} \rightarrow \phi_{-1} \mathrm{e}^{-i\alpha},\quad
\phi_{0}  \rightarrow   \phi_{0}\,.
\end{equation}
Remind that in the case of the hedgehog ansatz the iso-rotation of the configuration is equivalent to rotation of
the vector $\hat{r}_a$ by angle $\alpha$ around the $z$-axis.

The position curve of Hopfion is a is commonly chosen to be the curve $\phi^{-1}(0,0,-1)$, the preimage of the point $(0,0,-1)$ which is the antipodal to the vacuum $(0,0,1)$. For the simplest ${\cal A}_{1,1}$ Hopfion this is a circle of
radius $F(r_c) = \pi/2$, with numerical value $r_c = 0.8763$ in the $x-y$ plane. Small deviations $F(r) = F(r_0)+\epsilon$ then define the
tube around the position curve where $\vartheta \approx \pi/2$ and $\varphi = [0,2\pi)$.

From the parametrization \eqref{skyrmionAnsatzInphi} we can define the orientation of
the single Hopfion. Indeed, a clockwise rotation by an angle $\varphi$ in equatorial plane corresponds
to a counterclockwise rotation of the tube point on the target space
\begin{subequations}
\label{circle}
\begin{align}
\phi^1 \approx&  2\epsilon \sin \varphi,\label{cir1}\\
\phi^2 \approx& -2\epsilon \cos \varphi,\label{cir2}\\
\phi^3 \approx& -1.
\end{align}
\end{subequations}
Here, for the sake of convenience, we are using the component notations of the field
$\phi$.

Note that the Hopfion charge can be inverted
by the transformation $\BH \rightarrow \BH^* $ or $\phi_{a} \rightarrow (-1)^a \phi_{-a}$. It is easy to see that in this case
the signs of the right hand sides of \eqref{cir1} and \eqref{cir2} are inverted,
thus a clockwise rotation about the $z$ axis in the domain space corresponds to a
clockwise rotation on the target space. So for the Hopfion with negative topological charge
the point on the tube rotates in the opposite direction.
Thereafter we restrict our investigation to the case of positive values of $Q$ only.

For single Hopfion we can rotate the points on the tube by applying rotation transform via the $SU(2)$ matrix
\begin{equation}
\label{hopfionRotationM}
\BH\rightarrow  D(\alpha)\BH D(-\alpha)\, .
\end{equation}
Evidently this transformation is equivalent to \eqref{hopfionrotation} and it
leaves invariant the Lagrangian \eqref{modelInH}.

Let us now consider two identical Hopfions of degree one which are placed
at the points $\BR/2$ and  $-\BR/2$ and separated by a distance $R$, as shown in Fig.~\ref{fig:1}.
There the polar angle $\Theta$ yields the orientation of the Hopfions relative to the $z$-axis.
Note that in this frame the pattern of interaction is invariant  with respect to
the spacial rotations of the system around the $z$-axis by an azimuthal angle $\Phi$.

\begin{figure}[hbt]
\lbfig{fig:1}
\begin{center}
\includegraphics[height=.27\textheight, angle =0]{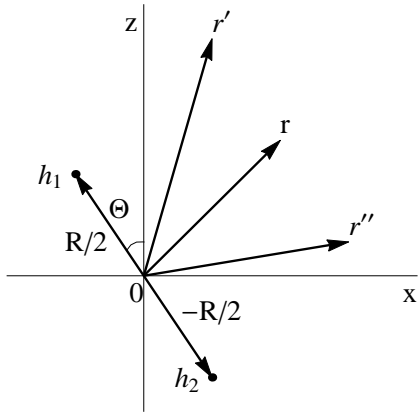}
\end{center}
\caption{\small Geometry of the system of two interacting Hopfions
$h_1$ and $h_2$,  placed at the points $\BR/2$ and
$-\BR/2$, respectively.}
\end{figure}

First, we suppose that both separated Hopfions are counterclockwise
oriented and they are in phase, i.e. $\Delta \alpha =0$. This system can be approximated by the product ansatz
\begin{equation}
\label{hopfionProductAnsatz}
\BH_2^{\Delta \alpha =0}(\Br)=2 \BU_0(\Br^\prime)\BU_0(\Br^{\prime\prime})\tau_0\BU_0^\dagger(\Br^{\prime\prime})\BU_0^\dagger(\Br^\prime)\,
\end{equation}
where
$\Br^\prime=\Br+\BR/2$ and $\Br^{\prime\prime}=\Br-\BR/2$.

Here fields of both Hopfions at the spacial boundary tend to the same
asymptotics $(0,0,1)$. Note, however, that in the constituent system \eqref{hopfionProductAnsatz} of two
identical Hopfions of degree one, contrary to the single Hopfion case,
the transformation \eqref{hopfionRotationM} of one of the Hopfions
$\BH$ do not leave the Lagrangian \eqref{modelInH} invariant, it
becomes a function of relative phase difference $\alpha$.

Further, in addition to the ansatz \eqref{hopfionProductAnsatz} we can consider
two separated Hopfions of degree one
with opposite phases, $\Delta \alpha =\pi$.
Using the definition \eqref{skyrmionAnsatz} we can express this system in terms of the matrix
$\BU_0(\Br)$, thus  the corresponding product ansatz is different from \eqref{hopfionProductAnsatz}:
\begin{equation}
\label{hopfionProductAnsatzAntiParallel}
\begin{split}
\BH_2^{\Delta \alpha =\pi}(\Br)=8
\BU_0(\Br^\prime)\tau_0\BU_0(\Br^{\prime\prime})\tau_0\BU_0^\dagger(\Br^{\prime\prime})\tau_0\BU_0^\dagger(\Br^\prime)\,,
\end{split}
\end{equation}

An advantage of the product ansatz approximations \eqref{hopfionProductAnsatz} and \eqref{hopfionProductAnsatzAntiParallel}
is that it ensures the conservation
of the total topological charge for any separation $R$ and space orientation of the constituents.
The simple additive ansatz of two unit charge Hopfions used by Ward \cite{Ward:2000qj} to construct the
Hopfion of degree two can be considered as a good approximation only if the Hopfions are well separated.

Substitution of product ansatzes \eqref{hopfionProductAnsatz} and \eqref{hopfionProductAnsatzAntiParallel}
into Lagrangian \eqref{modelInH} allows us to write down the expressions for the corresponding energy densities
of both  configurations as a function of the components of the position vectors
$r_i^{\prime}$ and $r_j^{\prime\prime}$ (cf Fig.~\ref{fig:1}).

Using the Gr\"{o}ebner basis method implemented in {\em Mathematica}, we can collect
these components into various combinations. It appears that in all cases the expressions  for the
local energy, as well as for the corresponding topological charge density
are some functions only of the distances $r^\prime$ and $r^{\prime\prime}$, the dot product
$\Br^{\prime}\cdot\Br^{\prime\prime}$, $z$-components of the vectors $r_0^{\prime}$, $r_0^{\prime\prime}$ and the
cross product $(\Br^{\prime}\times\Br^{\prime\prime})_0$.

Let us now express these quantities in terms of the Hopfion's position coordinates
$R,\Theta,\Phi$ and  the spherical coordinates $r,\theta,\varphi$, then the numerical integration of the corresponding
local densities  over the variables $\varphi,\vartheta$ and $r$ yields the total energy (mass) of the system and
its topological charge. In order to do it we apply some useful identities:
\begin{align}
\label{substitutions}
r^{\prime}=&\bigl(r^2+R^2/4+r R (\cos\Theta \cos\vartheta+\sin\Theta \sin\vartheta \cos(\varphi-\Phi))\bigr)^{1/2},\notag\\
r^{\prime\prime}=&\bigl(r^2+R^2/4-r R (\cos\Theta \cos\vartheta+\sin\Theta \sin\vartheta \cos(\varphi-\Phi))\bigr)^{1/2},\notag\\
(\Br^{\prime}\times\Br^{\prime\prime})_0=&\frac{r R \sin\Theta \sin\vartheta \sin(\Phi-\varphi)}{r^{\prime}r^{\prime\prime}},\notag\\
(\Br^{\prime}\cdot\Br^{\prime\prime})=&\frac{r^2-(R^2/4)}{r^{\prime} r^{\prime\prime}},\\
r^{\prime}_0=&\frac{r \cos\vartheta+(R/2)\cos \Theta}{r^\prime},\notag\\
r^{\prime\prime}_0=&\frac{r \cos\vartheta-(R/2)\cos \Theta}{r^{\prime\prime}}\,.\notag
\end{align}

Now, we illustrate the calculation procedure on a particular example of evaluation of
the local topological charge density, which in the circular coordinates is\footnote{Recall that the Hopf charge of
the configuration we constructed via projection
is given by the topological charge of the Skyrme field \cite{Battye1998}.}

\begin{equation}
\label{barDensGen}
\begin{split}
{\cal Q}(\Br^{\prime},\Br^{\prime\prime})=&\ii \sqrt{2}(-1)^{a+b} \left[
\begin{matrix}
1 & 1 &1\\
a & b & a+b
\end{matrix}
\right]
\mathop{\mathrm{Tr}}\Bigl(
\nabla_a \bigl(\BU_0(\Br^{\prime})\BU_0(\Br^{\prime\prime})\bigr) \BU_0^\dagger(\Br^{\prime\prime})\BU_0^\dagger(\Br^{\prime})\\
\times&\nabla_b \bigl(\BU_0(\Br^{\prime})\BU_0(\Br^{\prime\prime})\bigr) \BU_0^\dagger(\Br^{\prime\prime})\BU_0^\dagger(\Br^{\prime})
\nabla_{-a-b} \bigl(\BU_0(\Br^{\prime})\BU_0(\Br^{\prime\prime})\bigr) \BU_0^\dagger(\Br^{\prime\prime})\BU_0^\dagger(\Br^{\prime})
\Bigr)\,.
\end{split}
\end{equation}

Substitution of the product ansatz of two aligned ${\cal A}_{1,1}$ Hopfions
\eqref{hopfionProductAnsatz} into \eqref{barDensGen} yields the following expression for the topological charge density
\begin{equation}
\label{hopfionBarionDensityParallel}
\begin{split}
&{\cal Q}^{\Delta \alpha =0}(\Br^{\prime},\Br^{\prime\prime})=
-6\biggl( (1-(\Br^{\prime}\cdot\Br^{\prime\prime})^2)F^\prime(r^{\prime})
F^\prime(r^{\prime\prime})\Bigl(\frac{\sin (2 F(r^\prime))}{r^{\prime}}+\frac{\sin (2 F(r^{\prime\prime}))}{r^{\prime\prime}}\Bigr)
\\
&+\frac{2\sin ^2F(r^\prime)}{r^{\prime 2}} \Bigl(F^\prime(r^{\prime\prime})
   (\Br^{\prime}\cdot\Br^{\prime\prime})^2+F^\prime(r^\prime)\Bigr)
 +\frac{2\sin ^2F(r^{\prime\prime})}{r^{\prime\prime 2}} \Bigl(F^\prime(r^\prime)
   (\Br^{\prime}\cdot\Br^{\prime\prime})^2+F^\prime(r^{\prime\prime})\Bigr)\\
&+\frac{(1-(\Br^{\prime}\cdot\Br^{\prime\prime})^2)}{r^\prime r^{\prime\prime}}
\Bigl(\frac{\sin(2 F(r^\prime)) \sin ^2F(r^{\prime\prime})}{r^{\prime\prime}}\Bigr)
+\frac{\sin ^2F(r^\prime) \sin (2 F(r^{\prime\prime}))}{r^{\prime}}\\
&+\frac{2 \sin F(r^\prime) \sin F(r^{\prime\prime}) \bigl(F^\prime(r^\prime)+F^\prime(r^{\prime\prime})\bigr)}
{r^\prime r^{\prime\prime}} \Bigl(-2 \sin F(r^\prime) \sin F(r^{\prime\prime}) (\Br^{\prime}\cdot\Br^{\prime\prime})\\
&+ (1+(\Br^{\prime}\cdot\Br^{\prime\prime})^2) \cos F(r^\prime) \cos F(r^{\prime\prime})\Bigr)
\biggr)\,.
\end{split}
\end{equation}

It is possible to compute the total topological charge of the configuration to verify our construction for correctness.
This task becomes a little bit more simple since the expression \eqref{hopfionBarionDensityParallel} depends only
on the variables $r^\prime$, $r^{\prime\prime}$ and the dot product
$(\Br^{\prime}\cdot\Br^{\prime\prime})$. If we suppose that both Hopfions are sitting on top of each other, i.e.
$R\rightarrow 0$, then from \eqref{hopfionBarionDensityParallel} and \eqref{substitutions} we find
\begin{equation}
\label{parBarDensLimit}
\lim_{R\rightarrow 0} {\cal Q}^{\Delta \alpha =0}(\Br^{\prime},\Br^{\prime\prime})= -\frac{24 \sin^2\bigl(2F(r)\bigr) F^\prime(r) }{r^2}.
\end{equation}
Thus, this formula is different from its counterpart for the topological charge one configuration by factor of two.
Evidently, the total topological charge of the configuration then can be obtained by evaluation of the integral
\eqref{hopfionBarionDensityParallel} over the domain
\begin{equation}
Q=\frac{1}{24 \pi^2}\int_0^{2\pi}\dd\varphi\int_0^\pi\dd\vartheta\sin\vartheta\int_0^\infty\dd r r^2
{\cal Q}(\Br^{\prime},\Br^{\prime\prime})
\end{equation}
when the parameters $R$, $\Theta$ and $\Phi$ are arbitrary. Using the above mentioned
boundary conditions on the profile function $F(r)$, we arrived at $Q=2$, as expected.

The same procedure can be repeated when the Hopfions are in oposite phase, i.e. for the configuration given by the product anzats \eqref{hopfionProductAnsatzAntiParallel}.
In this case, however the corresponding topological charge
density depends on the variables $\Br_0^{\prime}$ and $\Br_0^{\prime\prime}$ as well as the above mentioned set of variables, thus
the result is a bit more complicated than \eqref{hopfionBarionDensityParallel} and is not repesented here.
Explicitly, in the limit $R\rightarrow 0$ it results in the function of the radial
variable $r$ and angle $\theta$ which is different from it counterpart \eqref{parBarDensLimit} and possesses a double zero at the origin
\begin{equation}
\label{opositeBarDensLimit}
\lim_{R\rightarrow 0} {\cal Q}^{ \Delta \alpha =\pi}(\Br^{\prime},\Br^{\prime\prime})= -\frac{96 \cos^2\theta \sin^4\bigl(F(r)\bigr) F^\prime(r) }{r^2}.
\end{equation}
However, the integration of this function subject of the same boundary conditions, also gives the same result $Q=2$ for any values of the
separation $R$ and orientation angles $\Theta$ and $\Phi$. Thus, we can identify this expression with
the topological charge density of the ${\cal A}_{2,1}$ Hopfion.

\section{Numerical results}

In a general case, evaluation of the total topological charge and the energy of the configuration constructed
via a product ansatz needs some numerical computations.
In Figs.~\ref{fig:4} and \ref{fig:5} the calculated isosurfaces of the topological charge densities are presented for some fixed
values of the set of orientation parameters, both for the Hopfions which are in phase and in the opposite phases,
i.e. applying the ansatz \eqref{hopfionProductAnsatz} and \eqref{hopfionProductAnsatzAntiParallel}, respectively.
Evidently, if the separation parameter $R$
is not very small, the product ansatz field given by \eqref{hopfionProductAnsatz},\eqref{hopfionProductAnsatzAntiParallel}
correctly reproduces the familiar structure of the system of two ${\cal A}_{1,1}$ Hopfions. Note that in the
first case, as separation parameter $R$ goes to some small but still non-zero value, the axially symmetric
charge two ${\cal A}_{1,2}$ unstable configuration of charge two \cite{Hietarinta:1998kt} is recovered.

\begin{figure}[hbt]
\lbfig{fig:4}
\setlength{\unitlength}{1cm}
\begin{center}
\hspace{0.5cm}
a)\includegraphics[height=.2\textheight, angle =0]{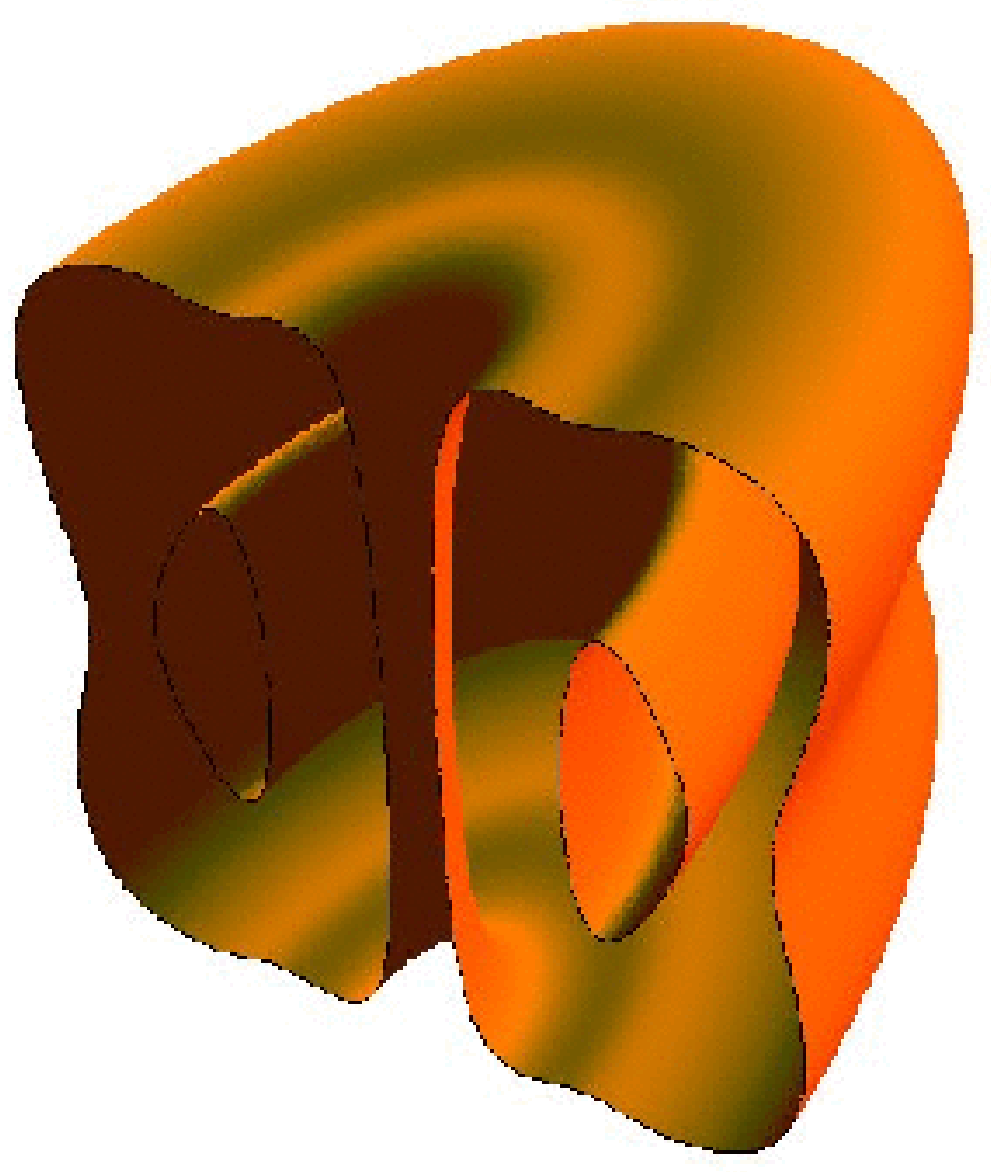}
b)\includegraphics[height=.2\textheight, angle =0]{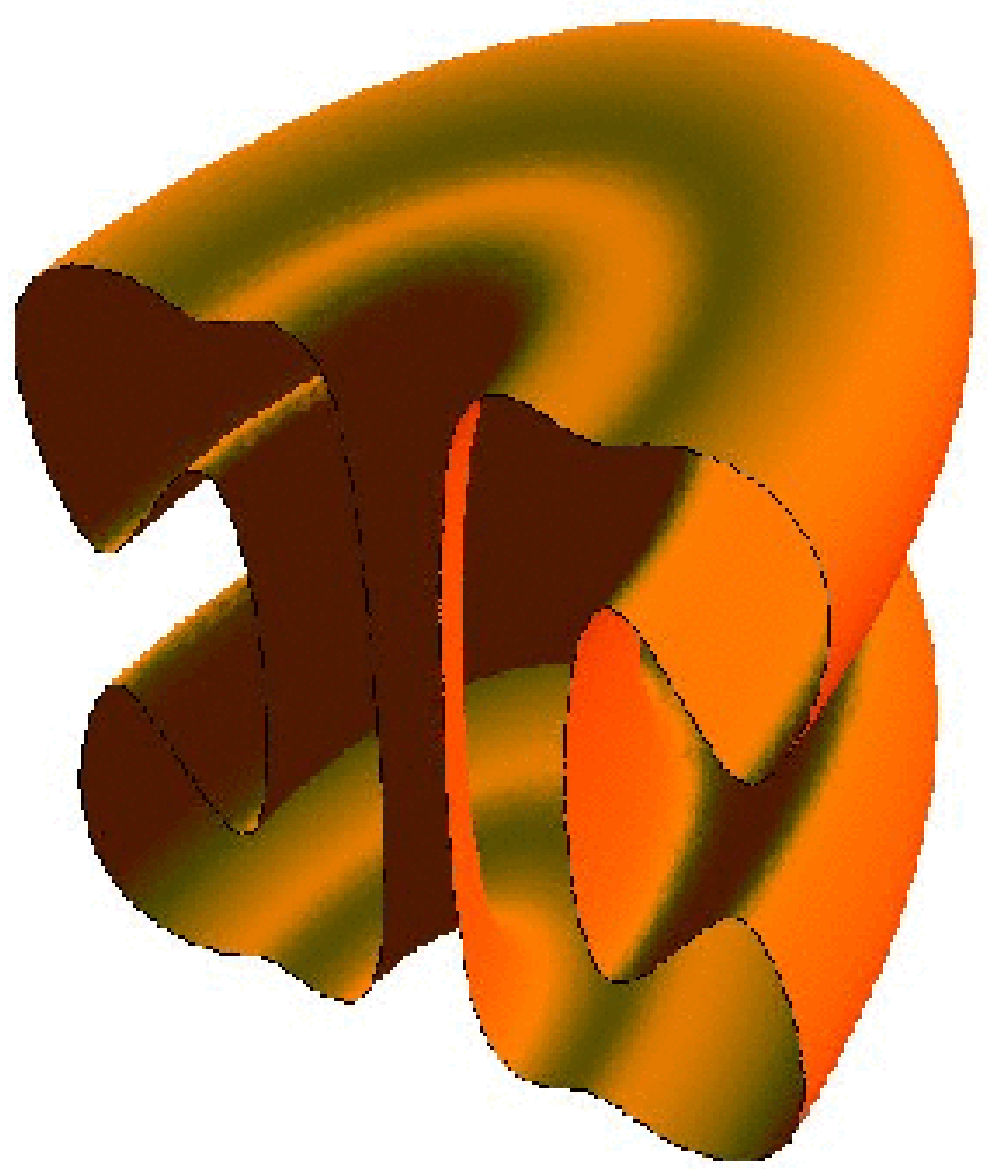}
c)\includegraphics[height=.23\textheight, angle =0]{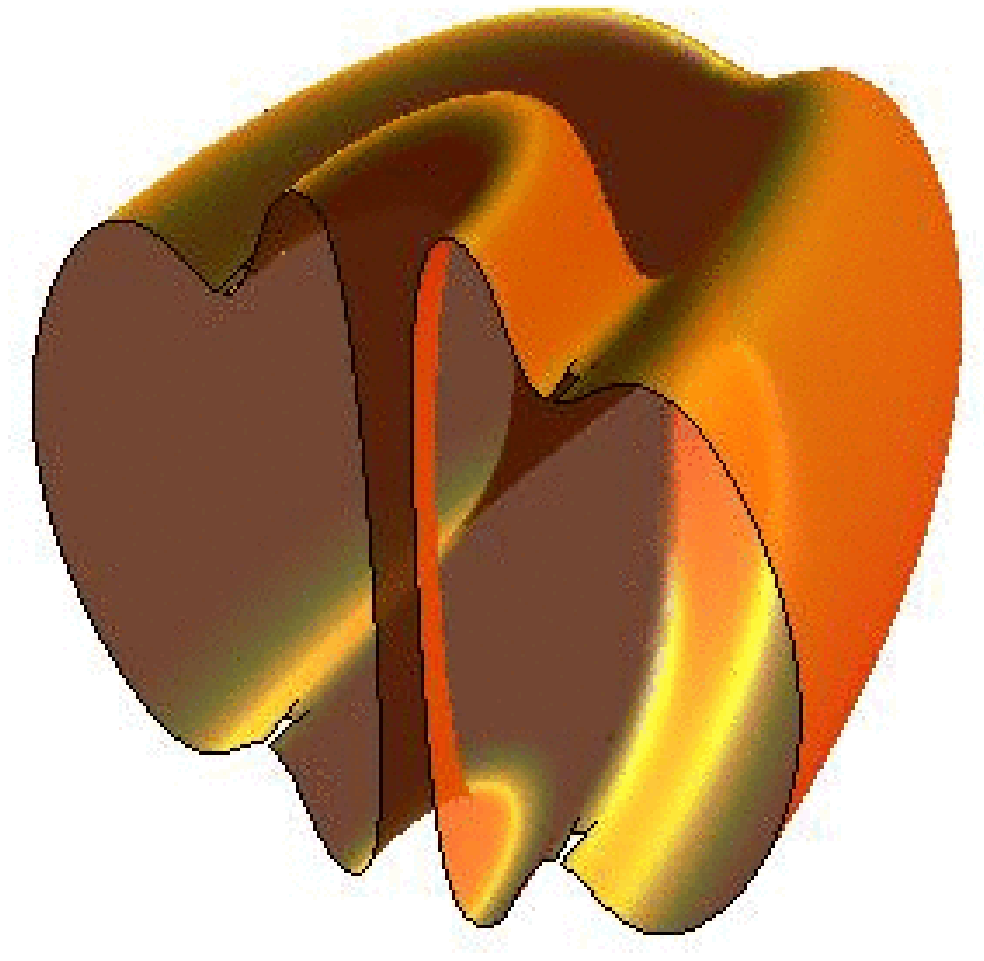}
d)\includegraphics[height=.23\textheight, angle =0]{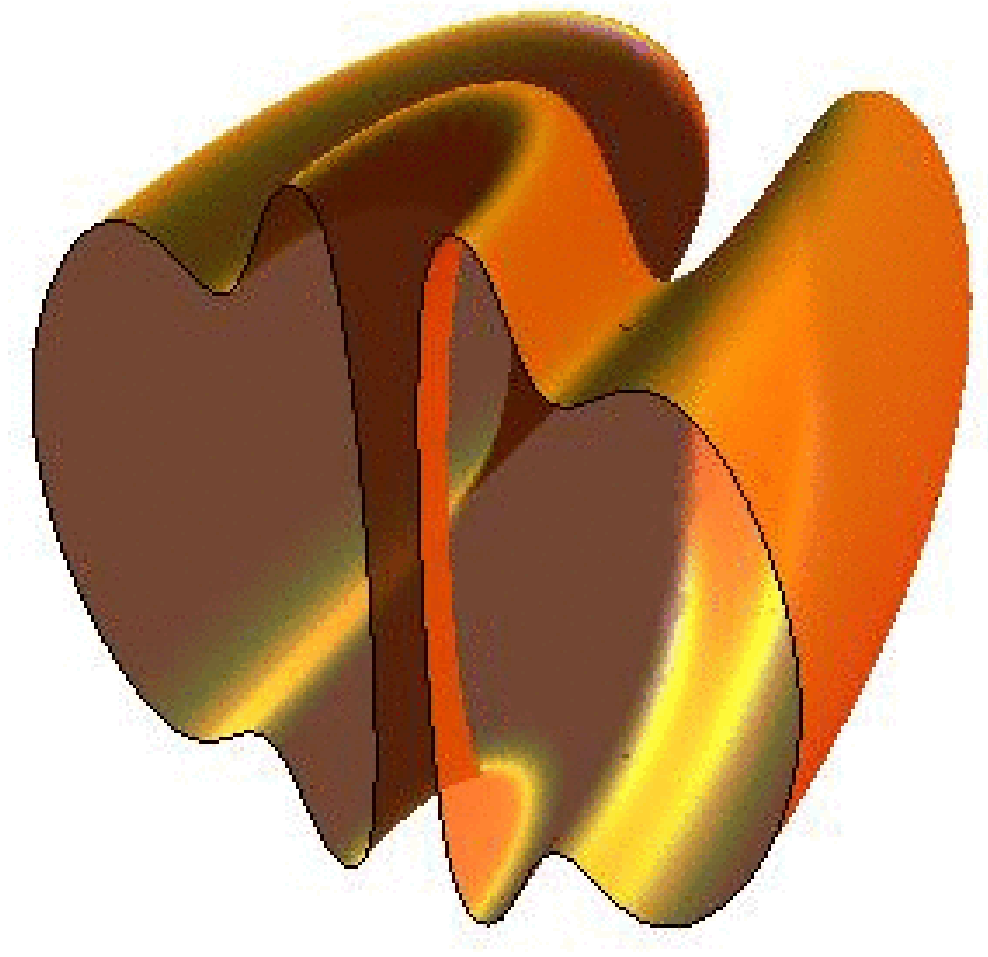}
\end{center}
\caption{\small Isosurfaces of the constant topological charge density of the product ansatz Hopfions which are
in phase, are presented at
${\cal Q}^{\Delta \alpha =0}(\Br^{\prime},\Br^{\prime\prime})=0.033$ (a)
and ${\cal Q}^{\Delta \alpha =0}(\Br^{\prime},\Br^{\prime\prime})=0.043$ (b)
for the set of orientation parameters $R=1,\Theta=0,\Phi=0$ (upper panel) and $R=1,\Theta=\pi/2,\Phi=0$ (plots (c),(d), bottom panel).
The surfaces are clipped through the vertical $\theta-r$ plane.}
\end{figure}

\begin{figure}[hbt]
\lbfig{fig:5}
\begin{center}
\hspace{0.5cm}
a)\includegraphics[height=.2\textheight, angle =0]{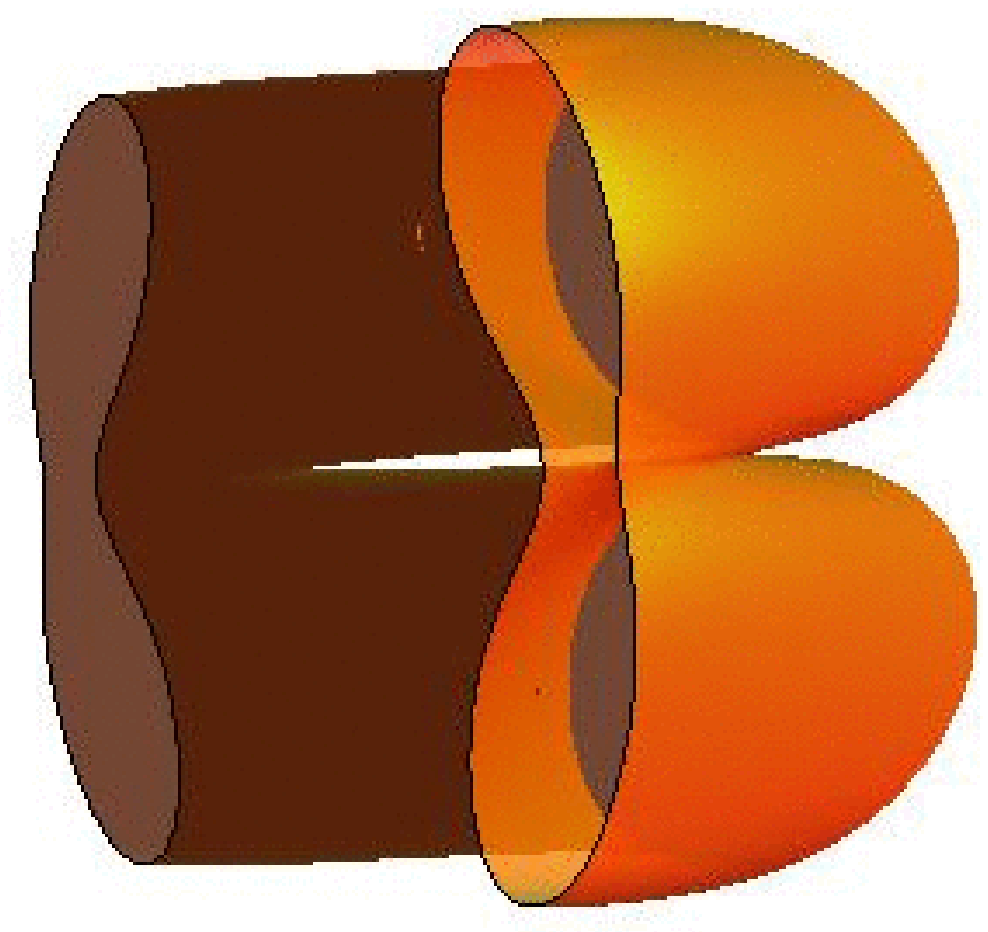}
b)\includegraphics[height=.2\textheight, angle =0]{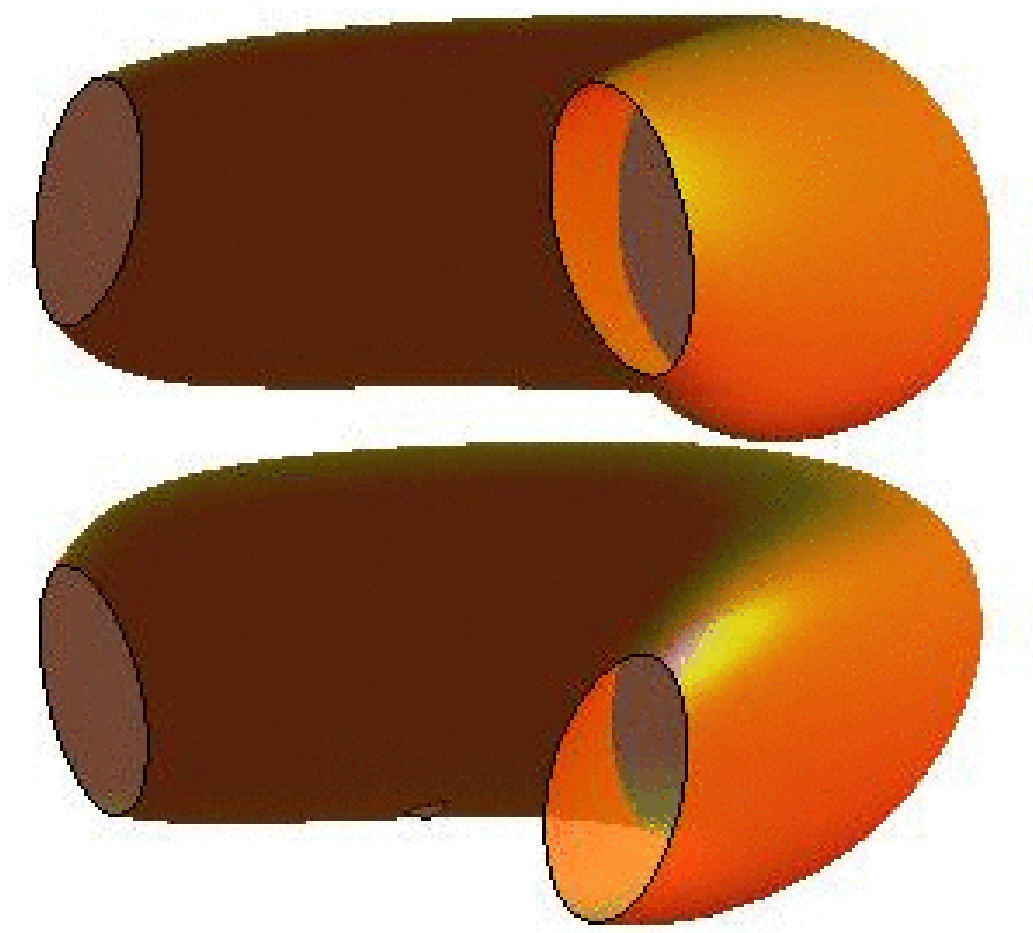}
c)\includegraphics[height=.2\textheight, angle =0]{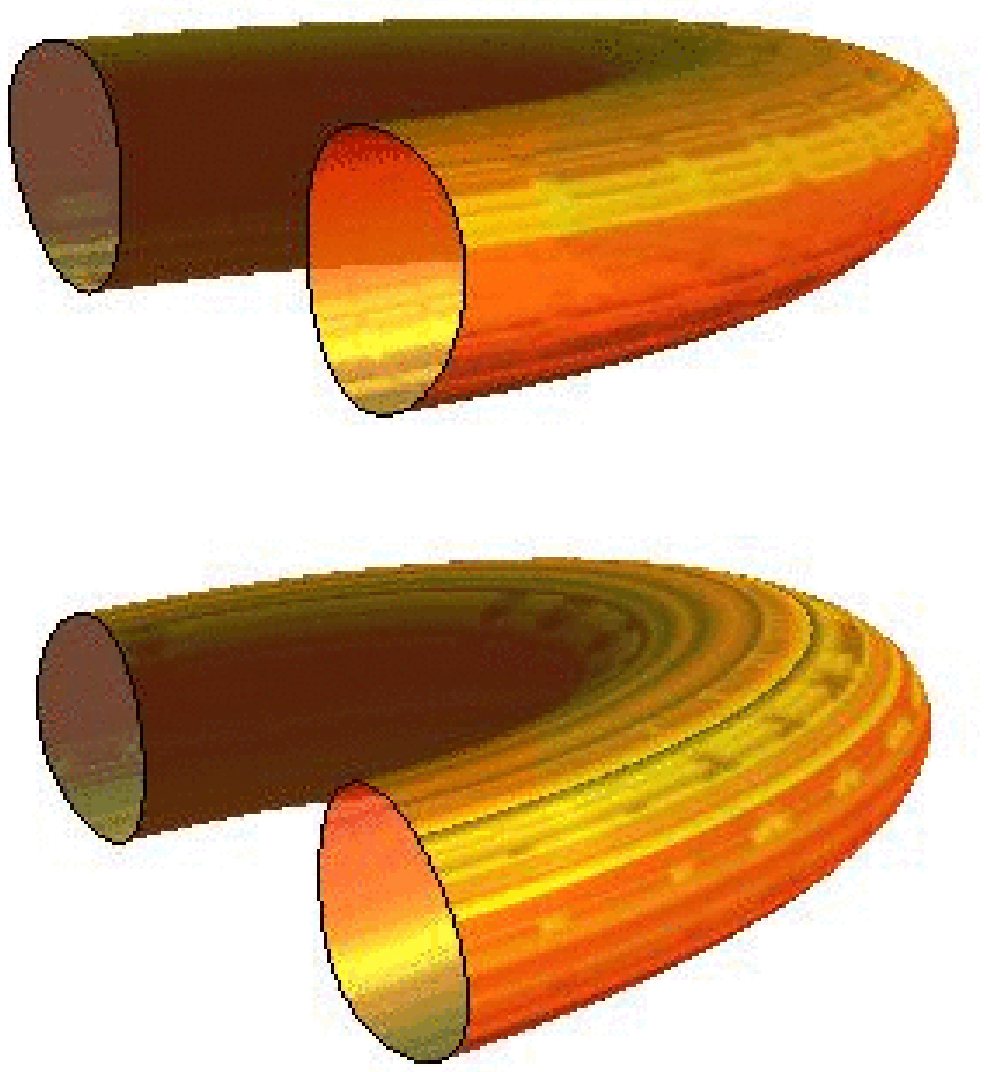}
\end{center}
\caption{\small
Isosurfaces of the constant topological charge density of the oposite phase product ansatz
Hopfions are presented for the set of orientation parameters
$R=1,\Theta=\pi/2,\Phi=0$ at ${\cal Q}^{\Delta \alpha =\pi}(\Br^{\prime},\Br^{\prime\prime})=0.113$ (a),
${\cal Q}^{\Delta \alpha =\pi}(\Br^{\prime},\Br^{\prime\prime})=0.133$ (b) and for the orientation parameters
$R=1,\Theta=0,\Phi=0$ at ${\cal Q}^{\Delta \alpha =\pi}(\Br^{\prime},\Br^{\prime\prime})=0.113$ (c).
The surfaces are clipped through the vertical $\theta-r$ plane.}
\end{figure}

Let us now evaluate the energy of interaction between the Hopfions.
Particularly, for each set of fixed values
of the orientation parameters $R$, $\Theta$ and $\Phi$, the integration of the energy density
yields the value of the interaction energy of the
Hopfions once the masses of two infinitely separated Hopfions (i.e. $M_0=2\times1.232\times 32\pi^2$)
are subtracted.

We have performed simulations with varying values of the parameters $R$, $\Theta$ and $\Phi$.
In Figs.~\ref{fig:2},\ref{fig:3} we presented the integrated product ansatz
interaction energy as a function of the orientation parameters
for in phase and oposite phase Hopfions, respectively\footnote{Note that in order to provide a reasonable approximation to
the system of two separated ${\cal A}_{1,1}$ Hopfions, the separation parameter $R$ must be larger that the size of the core $r_c$.}.

First, from our results we can conclude that the above product ansatz fields, both
\eqref{hopfionProductAnsatz} and \eqref{hopfionProductAnsatzAntiParallel}
correctly reproduce the pattern of interaction between the
Hopfions based on the simplified dipole-dipole approximation \cite{Ward:2000qj}. Indeed,
both for configurations which are in phase and in opposite phases,  the orientation along direction
given by $\Theta = 0$ matches the Channel A discussed by Ward \cite{Ward:2000qj}.
The orientation angle $\Theta = \pi/2$ corresponds to the Channel B.
Note that the Channel C represents the interaction between Hopfion and anti-Hopfion,
therefore is out of the scope of the present work.
Other values of $\Theta$ correspond to intermediate relative orientations of the Hopfions.

When the Hopfions are in phase and $\Theta = 0$ (Channel A) there is a shallow attractive window
for separations $R$ large than 4, as can be seen from Fig.~\ref{fig:2} (b). Evidently, this attractive channel
is very narrow because the potential of interaction
quickly becomes repulsive as the value of $\Theta$ increases. Note that the repulsive part of the potential
is concentrated inside the core where the product ansatz approximation is not very useful.

If $\Theta=\pi/2$ (i.e. the Hopfions are in side by side position),
the interaction potential is always repulsive as displayed in Fig.~\ref{fig:2} (c).
The energy of interaction for other orientations of the Hopfions is represented by a surface depicted in Fig.~\ref{fig:2}(a).

The pattern of interaction between the opposite phase ${\cal A}_{1,1}$ Hopfions is rather different.
Fig.~\ref{fig:3} (b) shows the corresponding energy of interaction as a function of the orientation parameters.
Evidently, in contrast with Fig.~\ref{fig:2} (b) in the Channel A ($\Theta = 0$ )
the interaction is always repulsive for any values of the separation parameter $R$.
However, in the Channel B ($\Theta = \pi/2$) the interaction energy is taking relatively large negative
value at the separation about the size of the core $r_c = 0.8763$ and then at gradually approach zero as separation between
the Hopfions increases, as shown in Fig.~\ref{fig:3} (c).

Finally, in Fig.~\ref{fig:3} (a) we depicted the energy of interaction between the Hopfions which are in opposite phases,
as function of the
orientation parameters $R$ and $\Theta$. We also have checked that the integrated interaction energy does not depend on the azimuthal
angle $\Phi$, as expected, though the expressions \eqref{substitutions} demonstrate that the energy density functional
explicitly depends on this orientation parameter\footnote{Mathematica notebook of all
calculation details can be downloaded from \url{http://mokslasplius.lt/files/Hopfion2013.tgz}.}.

To sum up, the product ansatz successfully captures the basic pattern of the interaction between the  ${\cal A}_{1,1}$ Hopfions,
our calculations suggest that for an arbitrary orientation of the Hopfions the system will evolve towards the state with minimal
energy shown in  Fig.~\ref{fig:3} (c). Qualitatively this conclusion is
in agreement with recent results of full 3d numerical simulations of the Hopfions dynamics
presented in \cite{Hietarinta:2011qk}.

\begin{figure}[hbt]
\lbfig{fig:2}
\begin{center}
a)\hspace{-0.6cm}
\includegraphics[height=.28\textheight,angle =0]{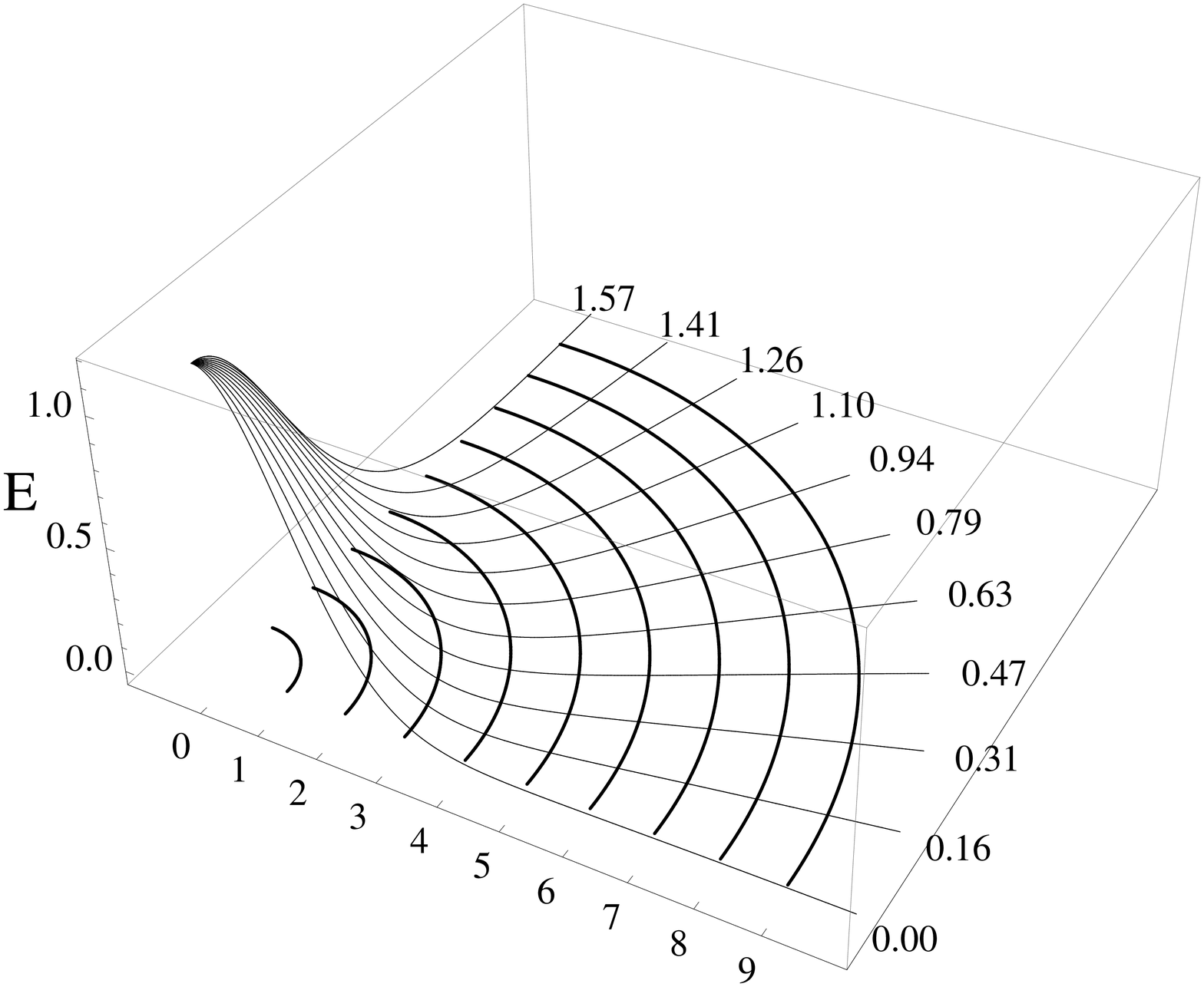}
\newline
b)\hspace{-0.6cm}
\includegraphics[height=.20\textheight, angle =0]{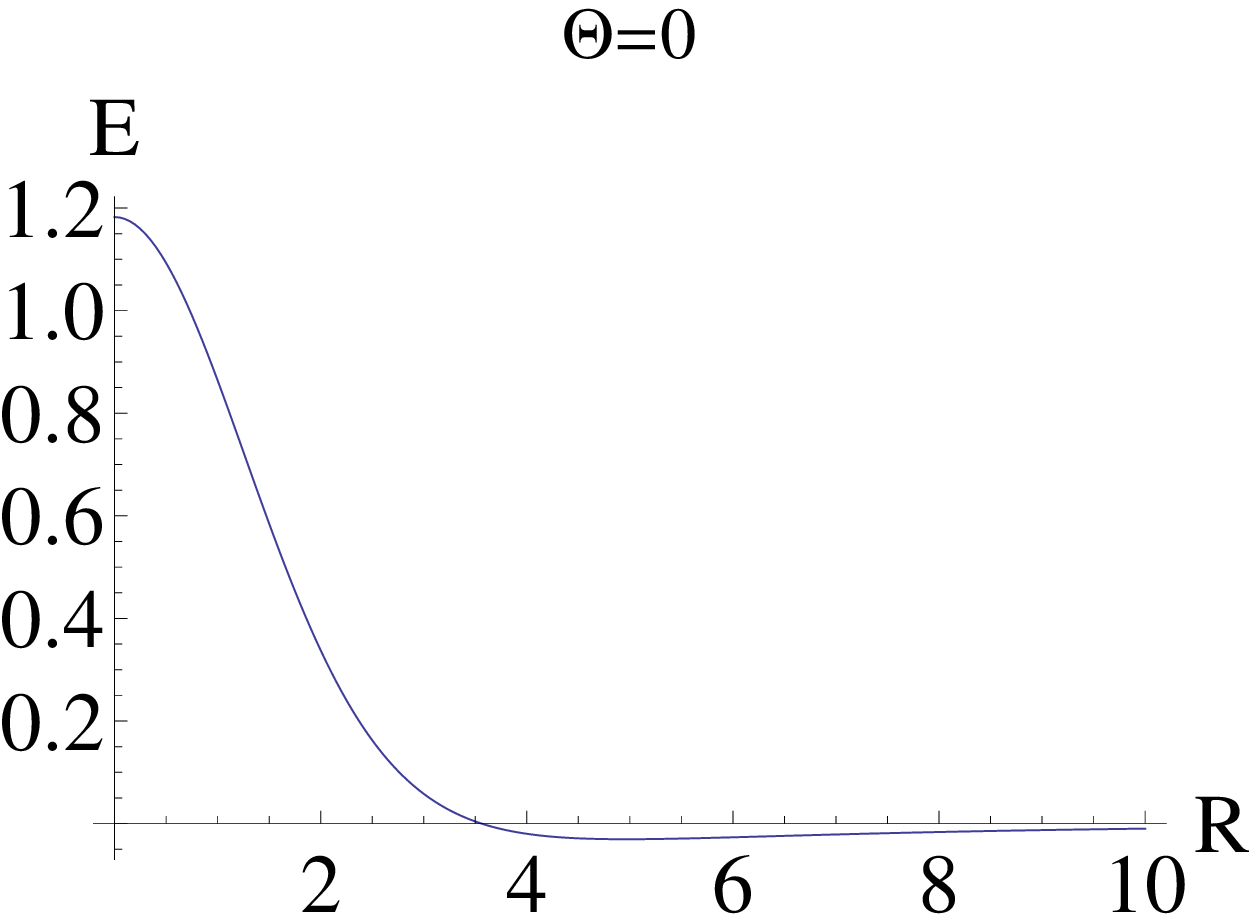}
c)\includegraphics[height=.20\textheight, angle =0]{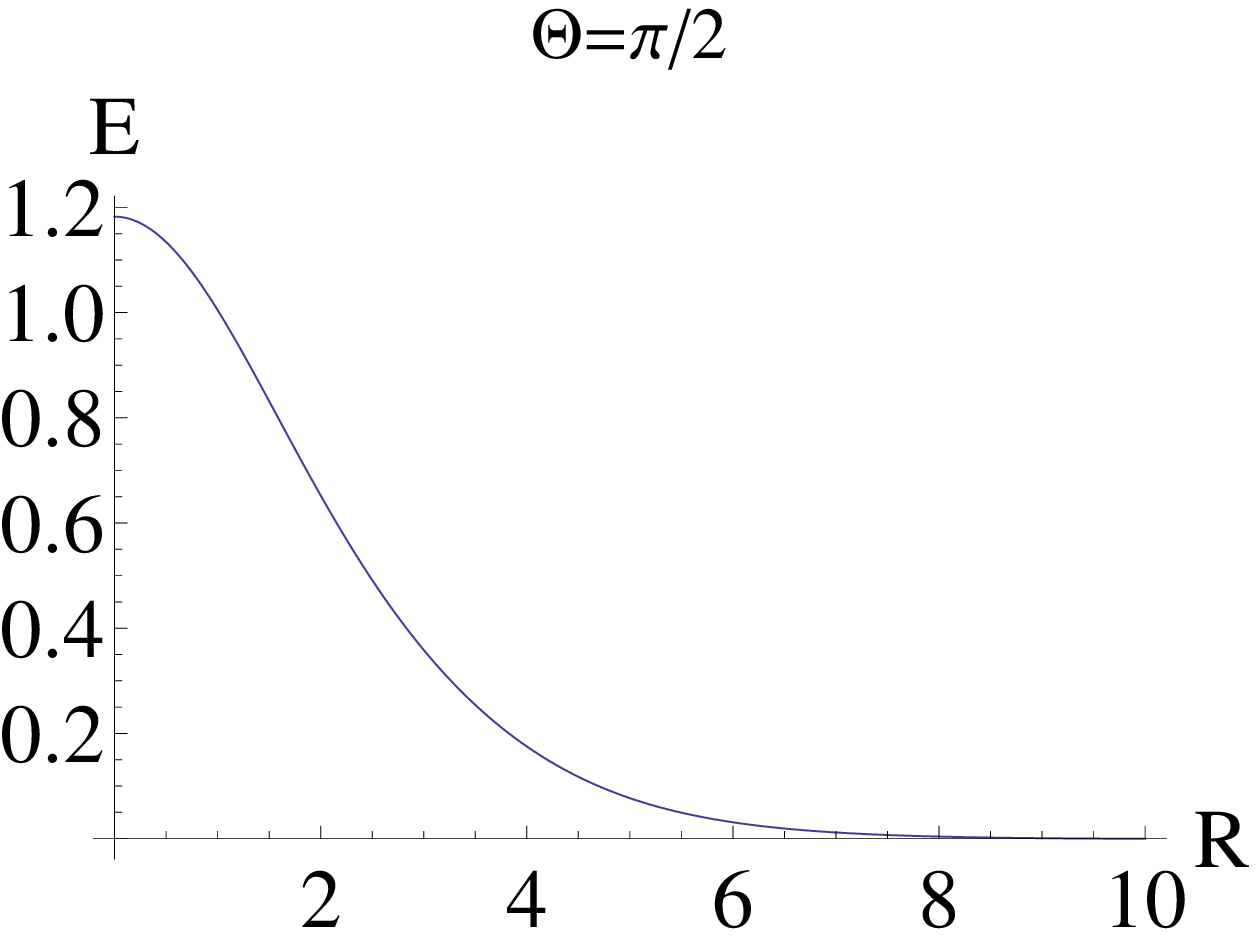}
\end{center}
\vspace{-0.5cm}
\caption{\small The evaluated interaction energy density of the  $\Delta \alpha =0$, product ansatz
Hopfions (in phase) as a function of the orientation parameters $R$ and $\Theta$.}
\end{figure}

\begin{figure}[hbt]
\lbfig{fig:3}
\begin{center}
\hspace{0.5cm} a)\hspace{-0.6cm}
\includegraphics[height=.28\textheight, angle =0]{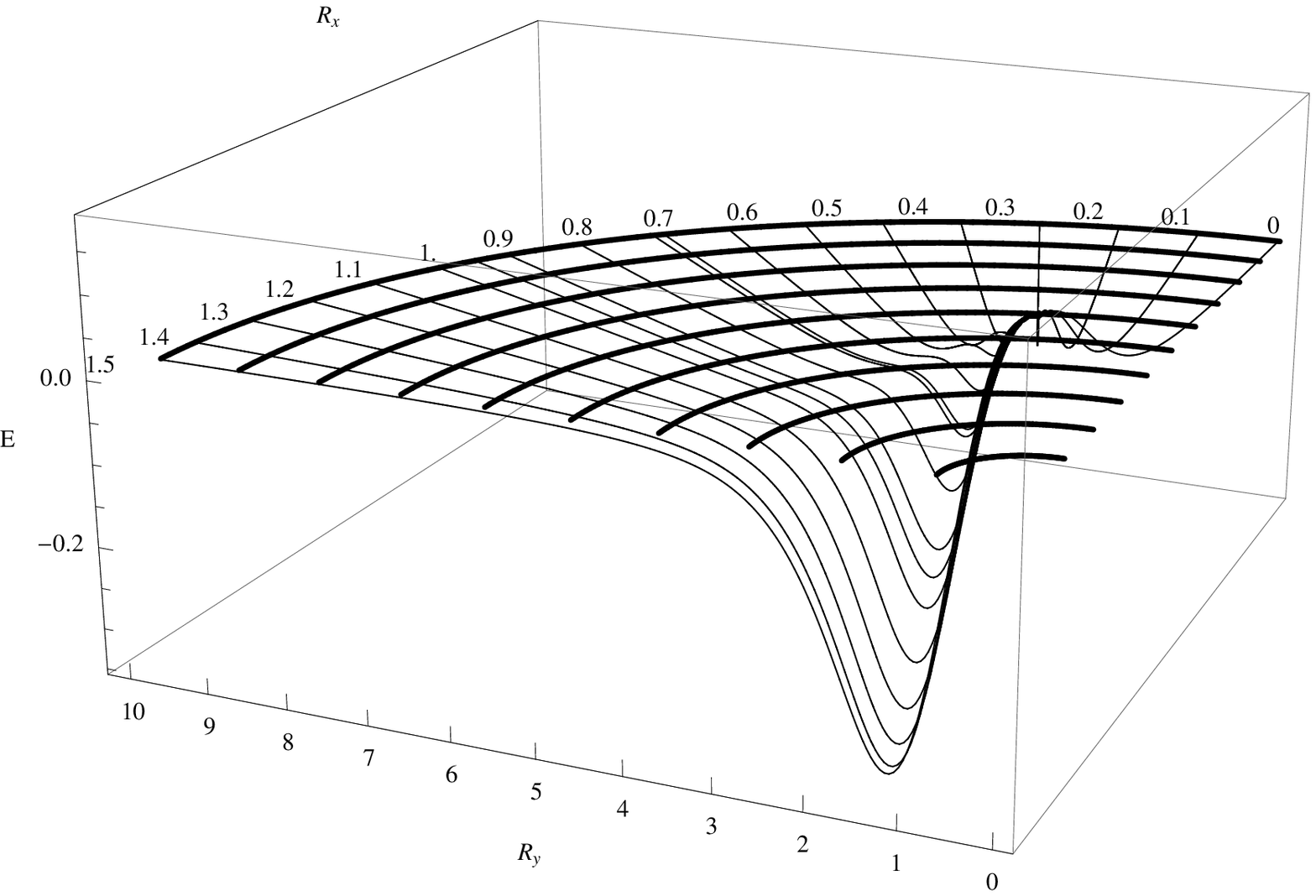}
\newline
b)\hspace{-0.6cm}
\includegraphics[height=.2\textheight, angle =0]{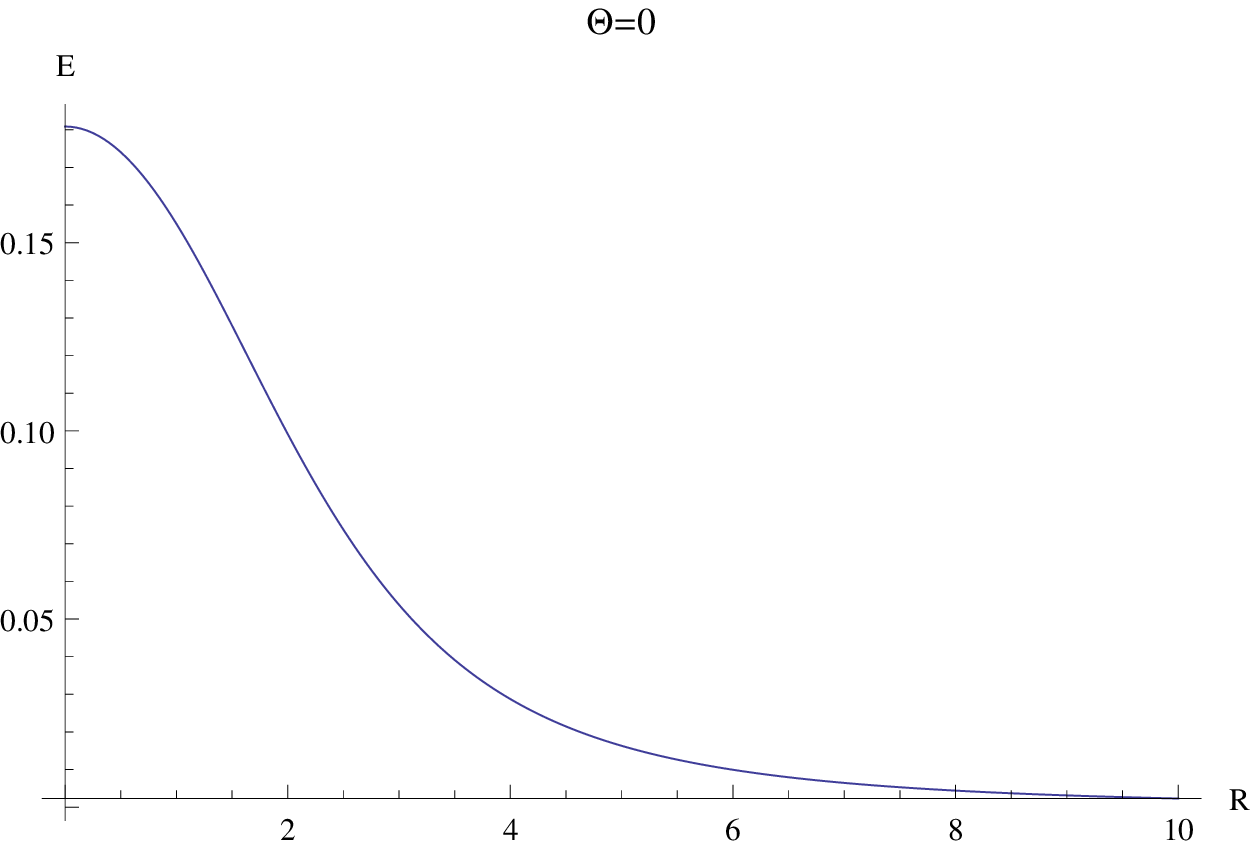}
c)\includegraphics[height=.2\textheight, angle =0]{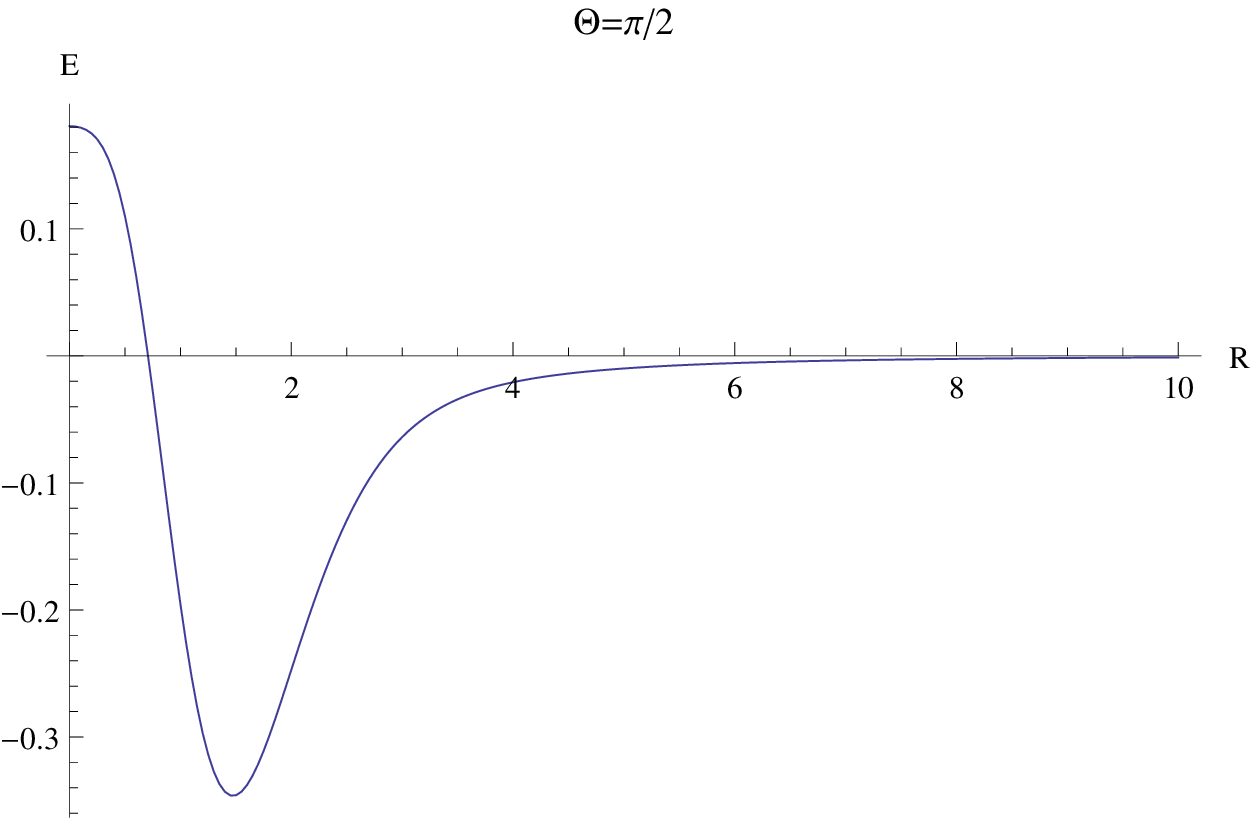}
\end{center}
\vspace{-0.5cm}
\caption{\small The interaction energy density of the  $\Delta \alpha =\pi$, product ansatz
Hopfions (opposite phases) as a function of the orientation parameters $R$ and $\Theta$.}
\end{figure}

\section*{Conclusion}

Using Hopf projection of the Skyrme field and the product ansatz approximation we
have investigated the pattern of interaction between the axially symmetric ${\cal A}_{1,1}$ Hopfions, in particular
we analysed how the interaction energy depends on the orientation parameters, the separation $R$ and the polar angle $\Theta$.
We have shown that this approach correctly reproduces both the repulsive and attractive
interaction channels discussed previously in the limit of the dipole-dipole interactions.
Here we mainly restricted our discussion to two most interesting cases considering in phase and oposite phase ${\cal A}_{1,1}$ Hopfions.

Finally, let us note that the product ansatz can be applied to construct a system of
interacting Hopfions of higher degrees. It can be done
if instead of the Skyrmion matrix valued hedgehog field \eqref{skyrmionAnsatz} we project the corresponding rational
map Skyrmions \cite{Houghton:1997kg,Battye1998}. On the other hand, setting the value of the separation parameter $R$
about the size of the core may be used to approximate various linked solitons, for example the configuration
${\cal L}_{1,1,1}^{1,1,1}$ can be constructed as a projection of the product
of three matrix valued Skyrme fields \eqref{skyrmionAnsatz}.


\section*{Acknowledgements}
This work is supported by the A.~von Humboldt Foundation (Ya.S.) and also from European
Social Fund under Global Grant measure, VP1-3.1-\v{S}MM-07-K-02-046 (A.A.).

\begin{small}

\end{small}

\end{document}